\documentclass{desyproc}
\begin{document}
\title{Particle physics @ $\sqrt{s_{pp}}\ \rm{ >}$ 50 TeV with the Pierre Auger Observatory}

\author{{\slshape Petr Travnicek$^1$ for the Pierre Auger Collaboration$^2$}\\[1ex]
$^1$Institute of Physics of the Academy of Sciences of the Czech Republic,\\ Na Slovance 2, CZ-182 21 Prague 8, Czech Republic\\
$^2$Observatorio Pierre Auger, San Mart{\'\i}n Norte 304, 5613 Malarg\"ue, Argentina,\\ {\tt http://www.auger.org/archive/authors\_2013\_06.html }}

\contribID{56}


\acronym{EDS'2009} 

\maketitle
\begin{abstract}
The Pierre Auger Observatory in Argentina provides the largest data
sample of the cosmic ray events with energy above 10$^{18}$ eV. These high
energy events can be used to test our understanding of the hadronic
interactions at energies beyond the reach of colliders and to probe the
basic properties of these interactions such as the inelastic cross-section
of proton-air collisions. The combination of an array of surface detectors and the fluorescence
telescopes of the Pierre Auger Observatory reduces significantly the
dependency of the shower energy estimation on MC simulations. Despite 
that, the interpretation of mass sensitive quantities such as the shower 
maximum in terms of chemical composition of cosmic rays still depends on 
the hadronic interaction models. This contribution describes the main 
results of the observatory concerning the chemical composition of the
cosmic rays and focuses on the problem of muon deficit in hadronic
interaction models and on the estimation of proton-air cross-section from
air-shower data.

\vspace{0.5cm}
{Comments: Presented at EDS Blois 2013 (arXiv:1309.5705)}

Report-no: EDSBlois/2013/56
\end{abstract}

\section{Introduction}

The Pierre Auger Observatory is situated in the Argentinian province Mendoza, close to the city of Malarg\"ue. It consists of a 3000 km$^2$ surface detector array and a set of 24(+3) fluorescence telescopes. The surface detector stations are water Cherenkov tanks each equipped with 3 photomultipliers measuring light induced by electromagnetic particles and muons. Six fluorescence telescopes occupy one fluorescence detector building. In total four of these buildings are located on the array border on small hills and thus overlook the interior of the
array measuring the longitudinal profile of electromagnetic shower \cite{FD}.
  Already in the year 2008 the Observatory was fully completed with successful
operation of all four fluorescence detector buildings and by fulfilling the original aim of 1600 deployed and working surface detector stations. During the next years several observatory enhancements were built such as 3 High Elevation Atmospheric Telescopes (HEAT) and AMIGA Infill both used to to lower the energy threshold below 10$^{18}$~eV.

The Pierre Auger Collaboration reports many results in the field of ultra-high energy cosmic rays (UHECR). The energy spectrum with the observation of the rapid flux suppression above 5$\times$10$^{19}$ eV \cite{spec}, various studies of expected anisotropy of ultra-high energy events (eg. \cite{anis})  or the estimates of the upper limits on the cosmic-ray photon and the diffuse neutrino fluxes \cite{phot,neutrino} are among the most important scientific outcomes of this project. In this contribution we rather focus on the results that are related to particle physics. Namely, we concentrate on the estimation of the proton-air cross-section, the connection between the analysis of chemical composition of cosmic rays (CR) and the hadronic interaction models. We focus also on the muon deficit in current hadronic interaction models.

\section{Proton-air cross-section}
The number of particles in an Extensive Air Shower (EAS) as a function of the atmospheric
slant depth (the amount of atmosphere traversed from its upper edge in g/cm$^2$)
is called the shower longitudinal profile. Most of the EAS energy is released
via the electromagnetic sub-shower. Therefore, as in the electromagnetic calorimeter 
the shower size increases until the average energy of the e$^\pm$ in the EAS 
is about the critical energy. The slant depth at which the longitudinal profile of a shower reaches
 its maximum is denoted as X$\rm{_{max}}$ and it is one of the primary observables measured by the fluorescence detector. Measurements of X$\rm{_{max}}$ can be used both to get insight into the composition of primary cosmic ray particles (showers induced by light particles penetrate deeper in the atmosphere than the showers originating from heavier nuclei) and also to measure the proton-air cross-section.  

\begin{figure}[hb]
\begin{center}
\includegraphics[width=0.43\textwidth]{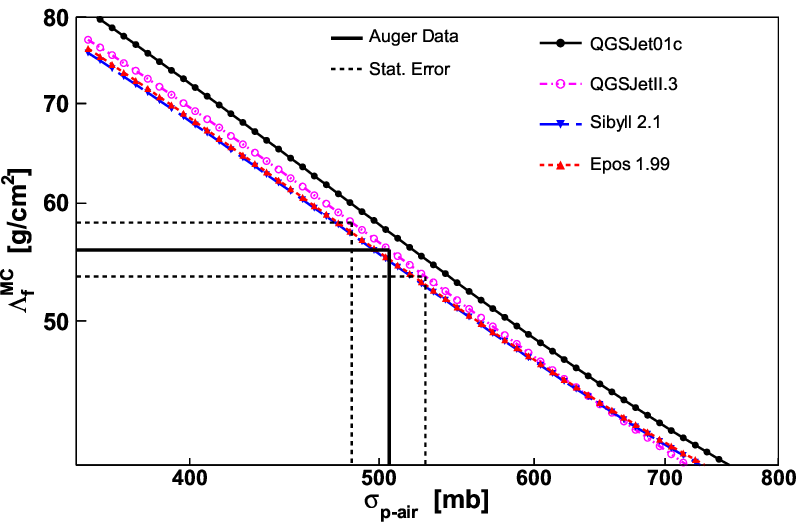}
\includegraphics[width=0.45\textwidth]{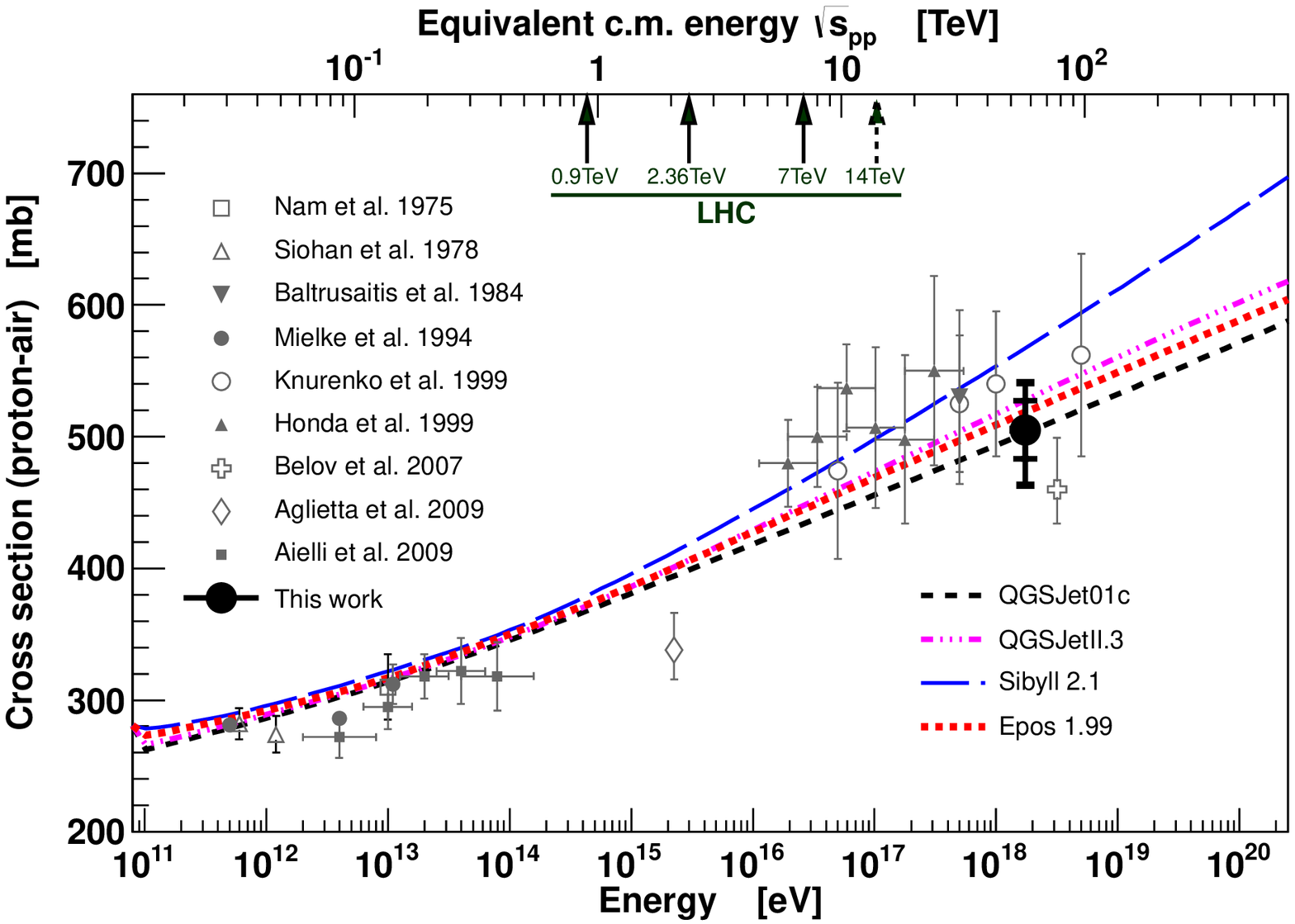}
\end{center}
\caption{Left: Dependency between the fitted slope ($\Lambda_f^{MC}$) of the X$_{max}$ distribution tail and the inelastic cross-section for several hadronic interaction models. The estimated slope is obtained by fitting the tail of the X$_{max}$ distributions  by  $dN/dX\rm{_{max}} \sim exp(-X\rm{_{max}}/\Lambda_f)$. Right: Resulting inelastic proton-air cross-section compared to other measurements and several hadronic interaction models.}\label{crossair}
\end{figure}

The differences in X$\rm{_{max}}$ between showers of the same primary energy and same primary particle type are due to fluctuations in hadronic interactions. For purely proton primaries, the X$\rm{_{max}}$ distribution is a convolution of the fluctuations in the shower development from the point of the first interaction to
the shower maximum (dependent on the hadronic interactions) and the exponential distribution of the depth of the first interaction. For this reason the fitted slope $\Lambda_f$ of the tail of the X$\rm{_{max}}$ distribution is (inversely) proportional to the inelastic cross-section of proton-air 
(see Fig.~\ref{crossair} - left).

 The data were selected in the energy range 10$^{18}$ $-$ 10$^{18.5}$ eV where the proton component is expected to dominate CR chemical composition. 
 To avoid biases in the measured X$\rm{_{max}}$ distribution fiducial volume cuts based on the shower geometry were applied. The largest source of systematic uncertainties is the lack of knowledge of the helium component. Uncertainties due to model assumptions were also addressed. For details see~\cite{crosssection}.
The value $\sigma_{p-air}^{inel} = 505 \pm 22 (\rm{stat}) ^{+28}_{-36} (\rm{syst})$~mb of inelastic proton-air cross-section was finally obtained at laboratory energy E$_{lab}$=10$^{18.24\pm0.005 (\rm{stat})}$ eV and compared to results of previous CR experiments (Fig.~\ref{crossair} - right).

\begin{figure}[hb]
\begin{center}
\includegraphics[width=0.45\textwidth]{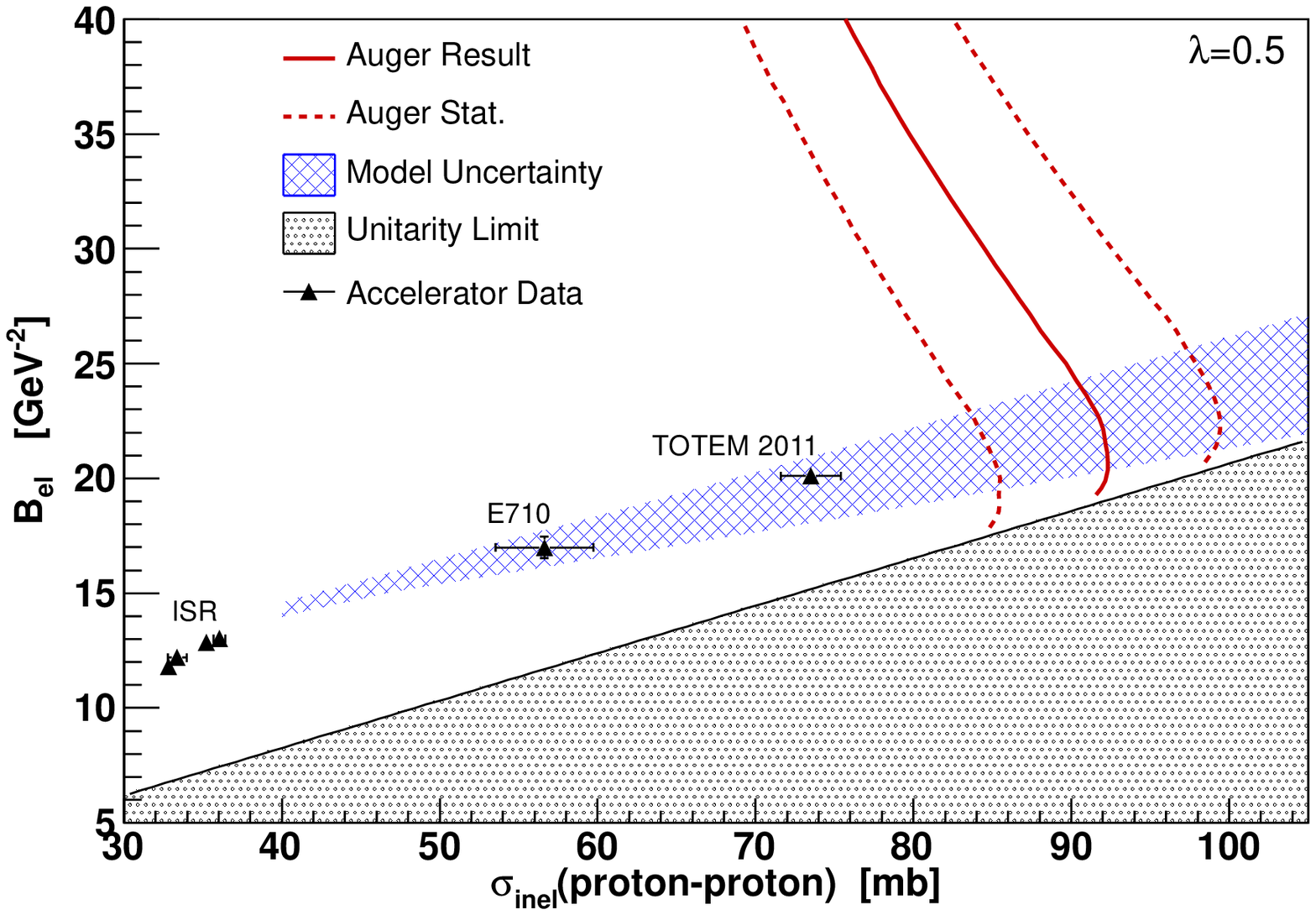}
\includegraphics[width=0.45\textwidth]{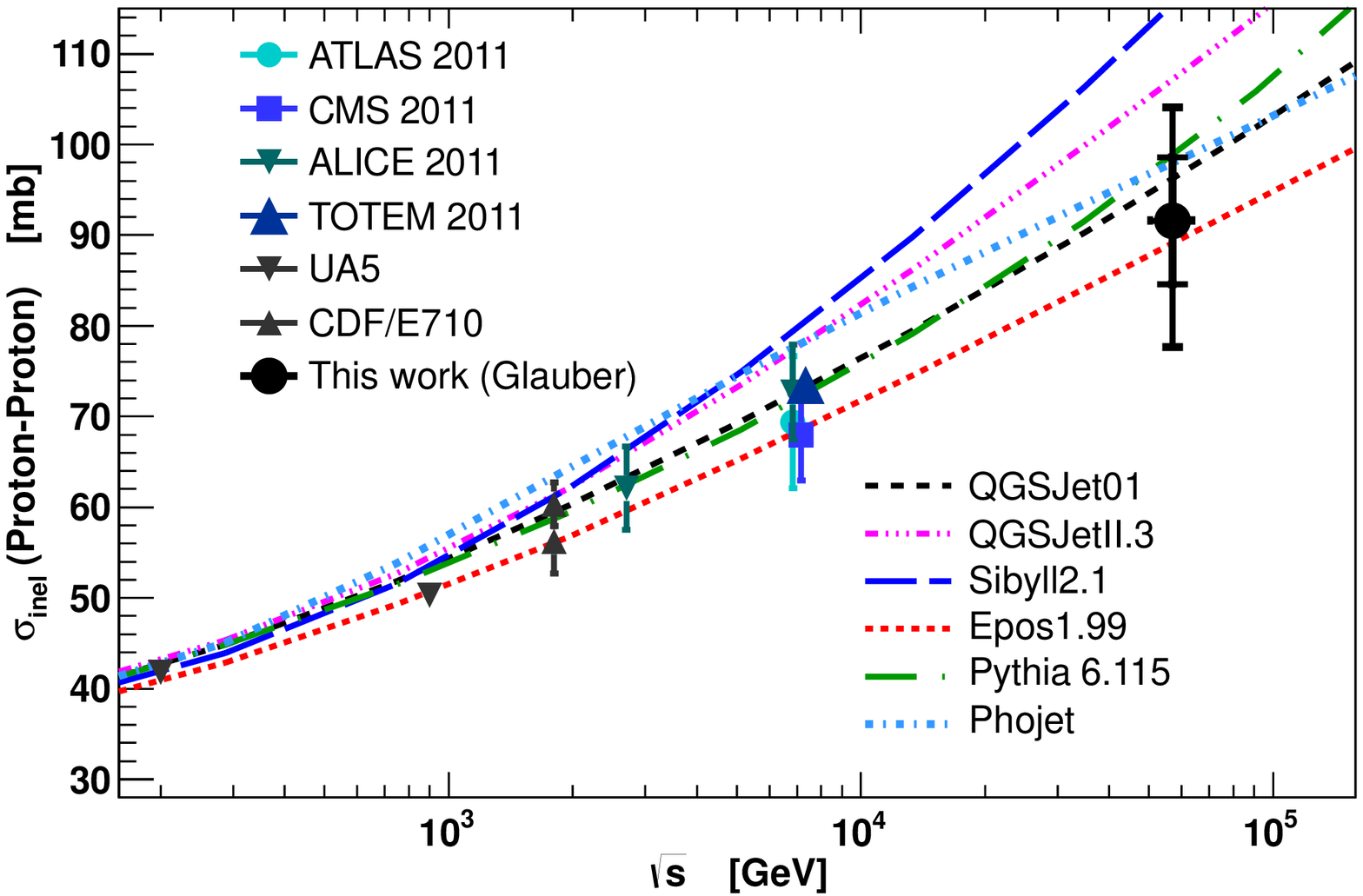}
\end{center}
\caption{Left: Correlation of elastic slope parameter,
$B_{el}$, and the inelastic proton-proton cross section in the Glauber
model. The solid line shows combinations of the parameters that
yield the observed proton-air production cross section, and
the dotted lines are the statistical uncertainties. The hatched area
corresponds to the predictions by SIBYLL, QGSJET, QGSJETII, and
EPOS.
Right: Derived $\sigma_{pp}^{inel}$  together with model predictions and accelerator data.}
\label{glaub}
\end{figure}

 Using the Glauber model (with intermediate inelastic states included) the proton-air cross-section can be converted to proton-proton inelastic cross-section ($\sigma_{pp}^{inel}$) at $\sqrt{s_{pp}}\ = 57 \pm 0.3 (\rm stat) \pm 3 (\rm syst)$~TeV. The conversion is illustrated in Fig.~\ref{glaub} (left) in the $\sigma_{pp}^{inel}$, $B\rm{_{el}}$ plane where B$_{el}$ is the elastic slope. The obtained value of  $\sigma_{pp}^{inel}$ is compared to results of LHC experiments and model predictions in Fig.~\ref{glaub} (right).  

\section{UHECR composition and hadronic interactions}
\begin{figure}[hb]
\begin{center}
\includegraphics[width=0.9\textwidth]{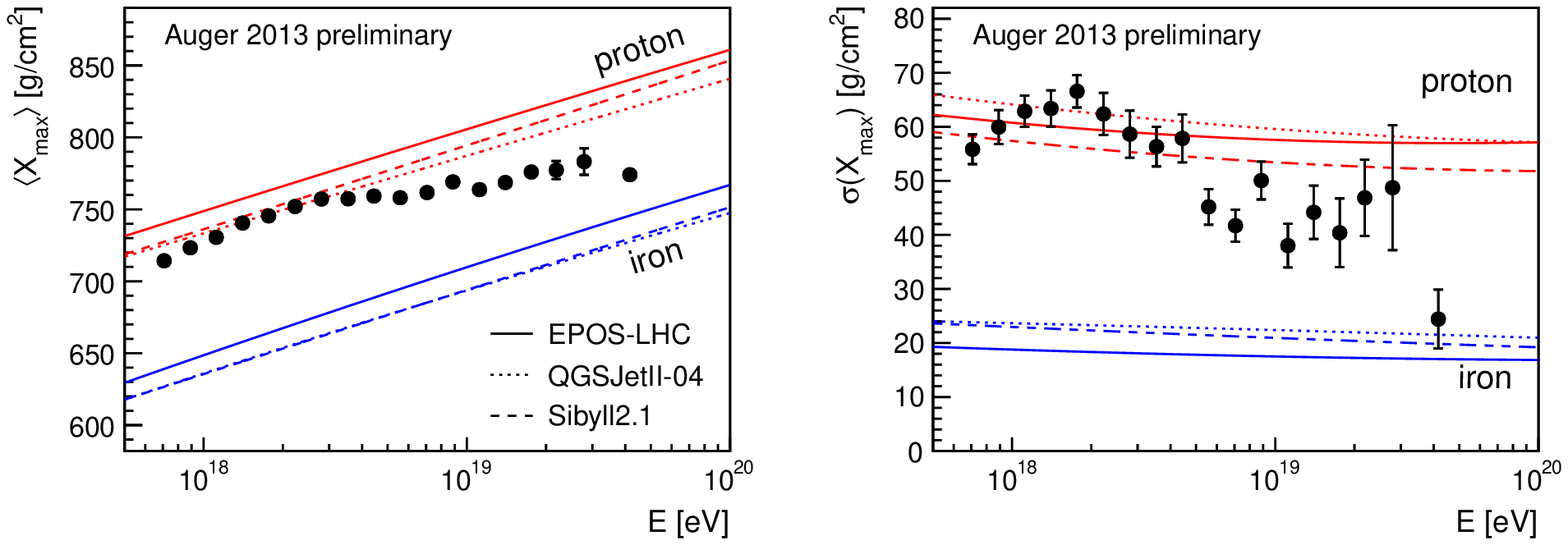}
\caption{Evolution of X$\rm{_{max}}$ (left) and $\sigma$(X$\rm{_{max}}$) (right) 
 as a function of energy~\cite{antoine}. Measurements are from the hybrid data set. Data (points) are shown with the predictions for protons and iron nuclei as primary particles for several hadronic interaction models.
}
\label{xmax}
\end{center}
\end{figure}

Quantitative estimates of the mass composition of cosmic rays deduced from 
measured mass sensitive observables can be obtained only when distributions of these quantities are compared to predictions of hadronic interaction models for different primary particle types. Hadronic interaction models, when used to describe UHECR showers, naturally rely on uncertain extrapolations to an order of magnitude larger centre-of-mass energies than are the energies currently achievable at accelerators. Even now when hadronic interaction models are extensively tested and tuned at the LHC it is not obvious at which point the models reliability would be such that they can be  universally used to unambiguously interpret UHECR shower data in terms of CR composition.
 For this reason the statements concerning the mass composition of cosmic rays have large uncertainties. The behavior of the 
X$\rm{_{max}}$ distributions (e.g. their main characteristics $\rm{<X\rm{_{max}}>}$ and $\sigma$(X$\rm{_{max}}$)) 
 strongly suggests that at energy $\sim$ 10$^{18.4}$ eV the mean mass of CR components starts to increase~\cite{xmaxprl,antoine} (see Fig.~\ref{xmax}). This conclusion is valid whatever model of hadronic interactions is compared to data of the Pierre Auger Observatory. In other words the only alternative explanation other than a sudden change of chemical composition deduced from the behavior of X$\rm{_{max}}$ is that the hadronic interactions change at that energy and that the important parameters inside the hadronic interaction models (such as the cross-section) must behave very differently than expected. 
The problems of the measurement of UHECR chemical composition and studies of hadronic interaction models are thus coupled subjects which shall be addressed at the same time. The cross-study of both problems using the first two moments of the X$\rm{_{max}}$ distribution ($\rm{<X_{max}>}$ and $\sigma$(X$\rm{_{max}}$)) is presented in \cite{petrera}.
 
\begin{figure}[hb]
\begin{center}
\includegraphics[height=0.32\textwidth]{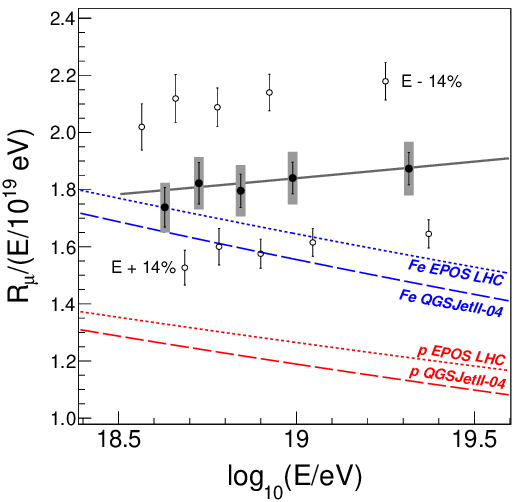}
\hspace{1cm}
\includegraphics[height=0.32\textwidth]{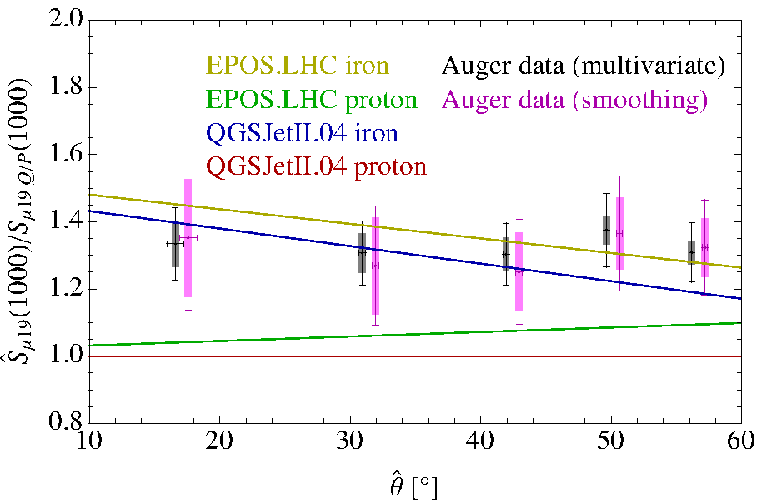}
\end{center}
\caption{Left:
Average value of muon number (scaled to E = 10$^{19}$ eV)
R$_\mu$/(E$_{FD}$/10$^{19}$ eV) relative to prediction of QGSJETII-03 (protons)
as a function
of shower energy. Theoretical curves for proton and
iron showers simulated with QGSJetTII-04 and EPOS-LHC
are shown for comparison. Open circles indicate the result
if the FD energy scale is varied by its systematic uncertainty.
 The gray thick error bars indicate the systematic uncertainty of
R$_\mu$.
Right:
The measured muon signal rescaling at E = 10$^{19}$ eV
and at 1000 m from the shower axis vs. zenith angle, with respect
to QGSJETII.04 proton as baseline. The rectangles represent the
systematic uncertainties, and the error bars represent the statistical
uncertainties added to the systematic uncertainties. The points for
Auger data are artificially shifted by $\pm$ 0.5 for visibility.}
\label{muons}
\end{figure}

 The situation gets even more complicated when other observables related to chemical composition are included. Namely, it was shown~\cite{icrc2013incl, icrc2013SD, icrc2013hyb} that current models of hadronic interactions
 predict fewer muons at ground than what is actually observed at the Pierre Auger Observatory.
 Figure~\ref{muons} shows an estimation of the relative number of muons in data normalized to predictions of QGSJETII-03 (Fig.~\ref{muons} left) and QGSJETII-04 (Fig.~\ref{muons} right) for proton primary particles at 10$^{19}$~eV. 
 The data are compared to predictions of two post-LHC interaction models QGSJETII-04 and EPOS-LHC for proton and iron nuclei as primary particles.
The figures summarize the results of independent studies, one done for inclined showers, with zenith angles larger than 62$^\circ$~\cite{icrc2013incl}, and another for showers with zenith smaller than 60$^\circ$~\cite{icrc2013SD} and using two different methods based on the time-structure of the surface detector signals to estimate the muon content of the shower.
The mentioned independent methods together with the analysis of hybrid events~\cite{icrc2013hyb} consistently indicate that the hadronic interaction models predict smaller size of the muon component than what is observed in the data unless pure iron composition is assumed. Such a heavy composition would be in contradiction with the X$\rm{_{max}}$ data when interpreted using the same models. This leads to the conclusion that shower models do not correctly describe the muonic ground signal.

\section{Conclusions}

The data from extensive air shower measurements can be used to address problems of hadronic interactions at energies far from the reach of current accelerators. Namely the inelastic proton-air cross-section at E$_{lab}$=10$^{18.24}$ eV was estimated and converted to inelastic p-p cross-section at  $\sqrt{s_{pp}} = 57$ TeV. The estimation of the chemical composition of cosmic rays and the behavior of the hadronic interactions at ultra-high energies are coupled problems. However, using any current model of hadronic interactions to interpret the evolution of the measured X$\rm{_{max}}$ distributions with energy, an increase of the mean mass of CR species is needed above $\sim$ 10$^{18.4}$ eV. A muon deficit in the predictions based on current hadronic interaction models
 is observed when compared to the data of the Pierre Auger Observatory.

\section{Acknowledgments}

The author thanks the organizers of EDS-Blois2013 and the colleagues from the Pierre Auger Observatory for opportunity to participate in the conference. This contribution was supported by the Ministry of Education, Youth and Sports of the Czech Republic within the project LG13007.


\begin{footnotesize}

\end{footnotesize}
\end{document}